\documentclass[journal]{IEEEtran}
\usepackage[T1]{fontenc}
\usepackage{multirow}
\usepackage{ifpdf}
\usepackage{subcaption}

\def\BibTeX{{\rm B\kern-.05em{\sc i\kern-.025em b}\kern-.08em
		T\kern-.1667em\lower.7ex\hbox{E}\kern-.125emX}}
	
\usepackage{cite}
\usepackage{amsmath,amssymb,amsfonts}
\usepackage{graphicx}
\usepackage{array}
\usepackage{fixltx2e}
\usepackage{stfloats}
\usepackage{pdfpages}
\usepackage{color}
\usepackage{soul}
\usepackage{caption}
\usepackage[export]{adjustbox}
\usepackage[T1]{fontenc}
\usepackage[english]{babel}
\begin{document}

\title{Transmit Antenna Selection for Massive MIMO-GSM with Machine Learning}

\author{Selen Gecgel, Caner Goztepe, Gunes Karabulut Kurt
	\IEEEauthorblockA{ \\{\{gecgel16, goztepe, gkurt\}}@itu.edu.tr}}

\maketitle

\begin{abstract}
A dynamic and flexible generalized spatial modulation (GSM) framework is proposed for massive MIMO systems. Our framework is leveraged on the utilization of machine learning methods for GSM in order to improve the error performance in presence of correlated channels and channel estimation errors. Both decision tree and multi-layer perceptrons approaches are adopted for the GSM transmitter. Simulation results indicate that in presence of real-life impairments machine learning based approaches provide a superior performance when compared to the classical  Euclidean distance based approach. The observations are validated through measurement results over the designed $16\times 4$ MIMO test-bed using software defined radio nodes.
\end{abstract}

\begin{IEEEkeywords}
Massive MIMO, Generalized Spatial Modulation, Machine Learning, Neural Networks, imperfect channels.
\end{IEEEkeywords}

\IEEEpeerreviewmaketitle
\section{Introduction}

\IEEEPARstart{T}{he} variability and the diversity of user expectations increase functionality requirements of wireless communication systems. In the fifth generation (5G) mobile broadband systems, it is aimed to improve the functionality of the systems as well as the capacity and the coverage. Massive multiple input multiple output (Ma-MIMO) technology is considered as one of the key techniques to address reliability, spectral and energy efficiency objectives of 5G and beyond systems\cite{S1}. 

Ma-MIMO systems with antenna selection (AS) methods provide a better performance by using the same number of RF chains when compared to conventional Ma-MIMO systems \cite{S2}. The authors of \cite{S12} proposed two transmit antenna selection methods: capacity-optimized antenna selection (COAS) and Euclidean distance-optimized antenna selection (EDAS). The EDAS method shows a better bit error rate (BER) performance with a higher complexity than the COAS method. Application of these methods in real-time systems is restricted by the complexity and flexibility requirements of next-generation communication systems. At this point, advances in machine learning (ML) techniques promise to leverage advantages of Ma-MIMO technologies. In  \cite{S4, S5}, AS problem is characterized as a multi-class classification problem, which enables to utilize ML based schemes. In  \cite{S4}, optimization-driven and data-driven AS methods are evaluated in terms of complexity and performance. In \cite{S5}, ML methods are implemented to select the optimal antenna that maximizes the secrecy performance. 

Generalized spatial modulation (GSM) is a practical candidate when the constraints of  Ma-MIMO systems are considered\cite{S6}. In \cite{S6}, a novel transmit antenna grouping scheme is proposed, improving the performance relative to the conventional GSM. In \cite{S7}, the authors show that GSM can achieve better performance than spatial multiplexing when the optimum antenna combination is selected. In \cite{S8}, it is shown that the channel estimation errors and correlated channels lead to a drop on the performance of GSM system.

In this letter, we propose to provide a more dynamic and flexible structure for Ma-MIMO systems. We propose to utilize ML methods for GSM to improve flexibility of Ma-MIMO systems and the contributions of the letter can be summarized as follows:
\begin{itemize}
	\item We firstly implemented GSM and also the applicability of machine learning methods is showed in the real-time environment.
	
	\item We utilize two ML based approaches, the decision tree (DT) and the multi-layer perceptron (MLP), for antenna subset selection in GSM under imperfect channel state information (CSI). 
	
	\item We analyze the performance of  proposed schemes numerically and experimentally, through the designed real-time software defined radio (SDR) based test-bed with $16 \times 4$ MIMO channel.
\end{itemize}

Both ML based approaches provide a better performance than the optimal AS method in terms of BER under non-ideal but realistic transmission scenarios. Simulations, as well as real-time test results, have shown that EDAS method is very successful when the channel is known perfectly and the signal to noise ratio (SNR) is high. Otherwise, our proposed schemes based on ML provide a superior error performance, which is demonstrated by simulations and real-time tests.

\section{System Model}
GSM can be seen as an extension of the spatial modulation (SM) system where each antenna index refers to different spatial bits so that more information is transmitted depending on the selected antenna\cite{pillay2017improved}. The proposed Ma-MIMO GSM system has ${N_t}$ transmit antennas and ${N_r}$ receiver antennas. 
Let the used number of antennas at the transmitter is $N_u$ and $I_A$ is the index symbol. The number of combinations that can be obtained with the GSM system and the maximum number of spatial bits that can be transmitted be denoted by $C = {{N_t}\choose{N_u}}$ and $ \lfloor \log_2 C \rfloor $ where $\left \lfloor . \right \rfloor$ denotes the floor operator, respectively. GSM eliminates the necessity of $N_t$ being power of $2$ and provides a more flexible usage of the system. The block diagram of the GSM system is shown in the Figure \ref{gsmblock}. The maximum number of bits that can be transmitted is $ \lfloor \log_2 M + \log_2 C \rfloor \ $ bits per channel use $(bpcu)$ where the modulation order is $M$. We assume that the communication channel is Rician, $\mathbf{H} \in {\mathbb{C}}^{N_r \times N_t}$ where $\mathbf{H^T}=[\mathbf{h}_1, \cdots,\mathbf{h}_i,\cdots, \mathbf{h}_{N_r}]$ and $\mathbf{h_i}=[h_1,h_2,\cdots, h_{N_t}]$. The received signal vector is \cite{S1}: 
\begin{equation}
\mathbf{y} = \mathbf{H}\mathbf{x} + \mathbf{w},
\end{equation}
here $\mathbf{w} \in \mathbb{C}^{N_r \times N_t}$ is an addictive white Gaussian noise. $\mathbf{x} = {s}_\zeta{\mathbf{I}^C}$ represents GSM symbol  where $s$ is a ${\zeta}^{\text{th}}$ $M$-QAM modulation symbol, $ \mathbb{S} $ is symbol cluster and $I^C_{t} \in \left \{  0, 1\right \}$ and $ \mathbf{{I}^C} = [I^C_1, I^C_2, \cdots, I^C_{N_t}]^\top \in \mathbb{I}$ is transmit antenna combination vector, $ \mathbb{I} $ denotes combination cluster. As a channel model, time-correlated and independent and identically distributed (\textit{i.i.d.}) are considered under perfect and imperfect channel estimates separately to represent the evolution from theoretical to practical systems. The imperfect estimate of the channel is as \cite{S9};
\begin{equation} \label{imp}
\widehat{\textbf{H}} = \sqrt{1-\beta}\textbf{H} + \sqrt{\beta} \textbf{e},
\end{equation}
where $\mathbf{H}$ is the perfect channel matrix. The time-correlated channel is modeled as follows\cite{S13};
\begin{equation} \label{corr}
\mathbf{{H_{\nu }}} = \sqrt{\alpha}\textbf{H}_{\nu -\iota} + \sqrt{1 -\alpha} \mathbf{z_\nu},
\end{equation}
where $\iota$ is the time offset. $\textbf{H}_{\nu }$ and  ${\mathbf{H}}_{\nu -\iota}$ represent the perfect channel matrix at time $\nu $ and ${\nu -\iota}$. The estimation error is represented by $\textbf{e}$, $\mathbf{z_\nu}$ is the time-correlation error and they are complex normal distributed with zero mean and unit variance. Recalling (\ref{imp}) and (\ref{corr}), the  imperfect time-correlated channel estimate is modeled as;
\begin{equation} \label{corrimp}
\widehat{\textbf{H}}_\nu  = \sqrt{(1-\beta )\alpha}\textbf{H}_{\nu -\iota} + \sqrt{(1-\alpha)(1-\beta)} \mathbf{z_\nu}  + \sqrt{\beta} \textbf{e},
\end{equation}
where $\alpha, \beta \in [0, 1]$, $\alpha$ is time correlation and $\beta$ is channel estimation error coefficients. 

A maximum likelihood detector is preferred at the receiver to identify data bits, in particular, spatial bits. The CSI must be accurately known for the best performance of the maximum likelihood detector. In GSM system designs, frequently it is assumed that CSI is known perfectly at the receiver, as the performance is improved under this assumption. The maximum likelihood detector is:
\begin{figure*}[!tb]
	\centering
	\includegraphics[width=15.5cm]{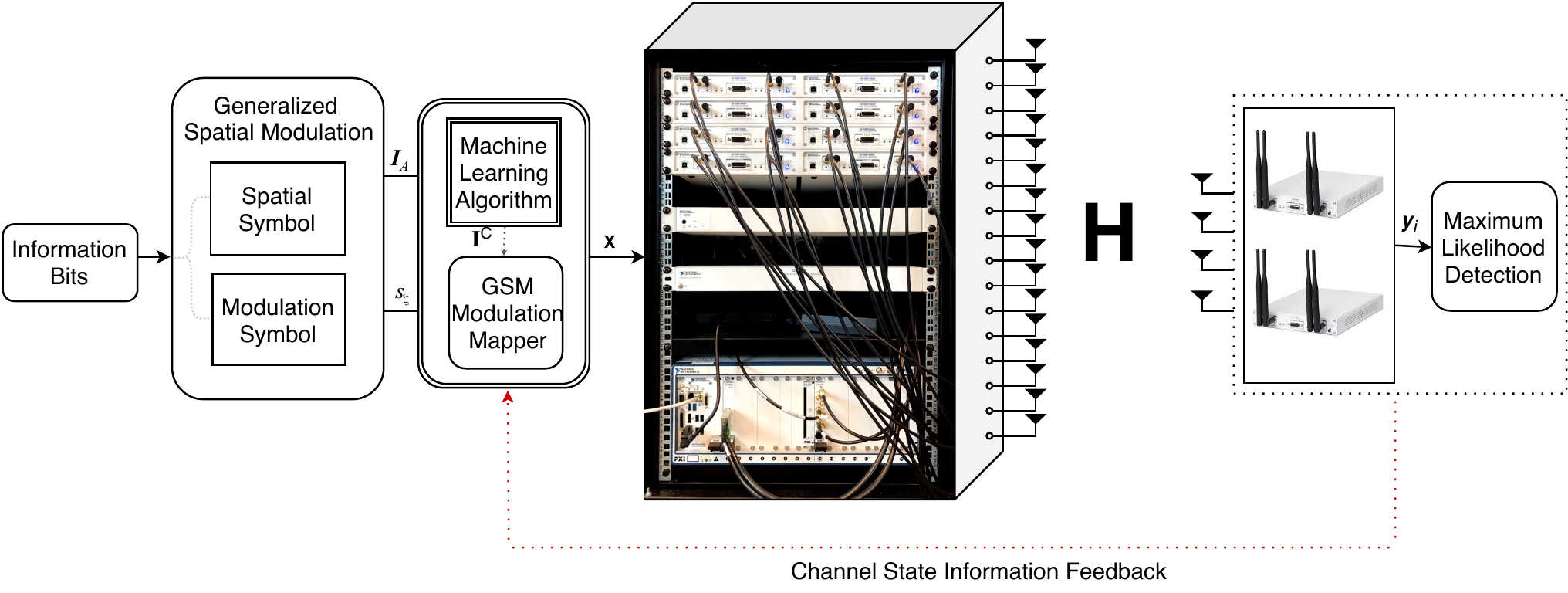}
	\caption{Proposed ML based antenna selection model for Ma-MIMO GSM} 
	\label{gsmblock}
\end{figure*}
\begin{equation}\label{eqn:gsm_ml}
(\hat{\mathbf{I}}^C, \hat{s}_{\zeta}) = \arg\underset{{\mathbf{I}}^C,\;{s}_{\zeta}}{\min} \sum_{i=1}^{N_r} {\left|{{y}_i} - s_\zeta  \mathbf{h_{i}}\mathbf{{I}^C}\right|}^{2}, 
\end{equation}
where ${y}_i$ is the received symbol at the $i^{\text{th}}$ receive antenna. The AS approach, which is frequently used in general MIMO systems, is seen as a suitable option for GSM-MIMO in terms of error performance. The optimal method for improving error performance is EDAS, which calculates the norm frame of the distance between the symbols under the perfect CSI. A subset $\kappa$ determined by EDAS is:
\begin{equation}\label{eqn:gsm_edasl}
\kappa_{EDAS} = \arg\underset{{\mathbf{I^C} \in \mathbb{I} }}{\max} 
\left \{{\underset{s_1 \neq s_2  \in \mathbb{S}}{\min} \ {\left | (s_1 - s_2 )\sum_{i=1}^{N_r}\mathbf{h_{i}} \mathbf{{I}^C}   \right | }}^{2}\right \}
\end{equation}

Although AS provides an improvement in error performance, it causes loss in spectral efficiency. Total subset number is $K = \lfloor \frac {C}{2^\varrho} \rfloor$ where $\varrho$ is the number of used spatial bits. In order to be able to perform AS, the $C$ in the GSM-MIMO system is divided into at least two subsets, causing a loss of at least 1 $bpcu$. Despite the increase in the number of groups and thus the loss of spectral efficiency, error performance is improved with AS.
\section{Proposed Scheme}
\subsection{Dataset Generation}
According to the model in Figure \ref{gsmblock}, four different cases are considered and four data sets  are built. The first data set is generated on real-time test-bed. To resemble the conditions of real-time systems, three different simulation based data sets are created depending on channel model equations (\ref{imp}), (\ref{corr}) and (\ref{corrimp}).  $\textbf{X} \ \epsilon \ \mathbb{R}^{N\times Q}$ and  $\textbf{X} = [x_1 \ x_2 \ ... \ x_{N-1} \ x_{N} ]^T$ represent one data set where $N$ is the number of samples in data set and $Q$ is the number of total features. Data sets are generated three attributes based on channel matrix and error vector magnitude (EVM) for each subset. An instance is designed as $\textbf{x}_n  = [f_{n,1} \ f_{n,2} \ ... \ f_{n,Q}]^T$. Three features are extracted for all subset as in below;\vspace{-2mm}
\begin{equation}
f_{n,q} =	\begin{cases}
\vspace{1mm}
\underset{r=1}{\overset{N_u}{\sum}} \ \underset{t=1}{\overset{N_t}{\sum}} \ \left | h_{t,r} \right | ^{2}, & \text{ if } q=1,\ \cdots \ ,K \\ 
\vspace{1mm}
\frac{\frac{1}{N} \underset{n=0}{\overset{N-1}{\sum}} \varepsilon} {P_S} \times 100  \ , & \text{ if } q=(K+1), \ \cdots \ , 2K \\ 
\textbf{H}_{n} - \textbf{H}_{n-\delta } \ , & \text{ if } q=(2K+1),\ \cdots \ , Q 
\end{cases} 
\end{equation}
where $K$ is the total subset number, $P_S$ is the average symbol power, $\varepsilon$ is the error power, $\delta$ is sample delay. Each class represents an antenna subset set to transmit a spatial symbol. In the labeling part of data set preparation, EDAS, optimal method in terms of error performance is used depending on the equation (\ref{eqn:gsm_edasl}).
\vspace{-2mm}
\subsection{Learning Algorithms}

\vspace{-0.7 mm}
\subsubsection{Decision Tree}
Decision tree algorithm is a non-parametric supervised learning method used for classification and regression. To estimate correctly the class of the new sample is the main objective by creating decision rules based on statistical information of the training data set \cite{S10}. The decision trees are composed of 3 basic units as root, branch and leaf nodes which form the structure of tree from the beginning to last. The attributes who maximizes information gain on each node is determined using the following equations and the node is divided into child nodes depending on the selected attributes.
\begin{equation}
\Upsilon_{m}= -\sum_{\kappa =1}^{K}p_{m}^{\kappa } \log_2 p_{m}^{\kappa} \ ,
\end{equation} here ${p_{m}^{\kappa} = N_{m}^{\kappa} / N_{m} }$ where $p_m ^\kappa$ represents the probability of the $\kappa^{\text{th}}$ class on the $m^{\text{th}}$ node, $N_m$ is the number of data samples, $N_m^\kappa$ shows the probability of samples belongs to $\kappa^{\text{th}}$ class  on $m^{\text{th}}$ node, $\Upsilon_{m}$ is the entropy of the $m^{\text{th}}$ node. On the each node for determining the feature which provides maximum information gain, \vspace{-1 mm}
\begin{equation}\label{gain}
\Psi(m, \textbf{f}_{q}) = \mathit{\Upsilon}_m -\sum_{d=1}^{D} p_{d}\mathit{\Upsilon}_d
\end{equation}
is used, where $ \textbf{f}_{q}$ is $q^{\text{th}}$ feature, $\Psi$ is information gain related to $ \textbf{f}_{q}$ on the $m^{\text{th}}$ node, $d$ shows the number of child nodes on the $m^{\text{th}}$ node, $p_d$ is the probability of the corresponding child node and $\mathit{\Upsilon}_d$ is the entropy of the $d^{\text{th}}$ node.
\subsubsection{Multi-Layer Perceptrons}
MLP is an artificial neural network structure and has the ability to learn non-linear functions via a given set of features. MLP is a considerably successful method, even though it has a higher computational complexity compared to decision tree. It consists of 3 main layers; input layer, hidden layers and output layer. The MLPs are trained with ADAM optimizer using backpropagation algorithms through neurons at these layers. On the training of the model, cross entropy $J(\varTheta)$ is used as a cost function as given in \cite{S11}.

\begin{equation} \label{eq:cross_entropy}
J(\varTheta)= \frac{-1}{N} {\sum_{n=1}^{N} \sum_{\kappa=1}^{K} y_\kappa^{(n)} \text{log}\rho_\kappa^{(n)}},
\end{equation} 
where $N$ is the number of instances and $\rho_\kappa^{(n)}$ is the estimated probability that the instance $x^{(n)}$ belongs to class $\kappa$. $y_\kappa^{(n)}$ is equal to 1 if the target class for the $n^{\text{th}}$ instance is $\kappa$. Otherwise, it is equal to 0. The hyperparameters of the training phase are determined empirically.

\section{Test-Bed Design}
The test-bed is designed as shown in Figure \ref{gsmblock}. In the proposed test system, eight USRP-2943Rs at the transmitter and two USRP-2940 (Universal Software Radio Peripheral) devices are used at the receiver. Each one is used as an SDR node with two radio frequency (RF) chains. As the software, LabVIEW, a graphical programming language is preferred. The PXIe-6674T module, which generates a reference clock, and the CDA-2990 module that share this clock information to all devices are used to provide hardware synchronization of the 12 transmitter channels in the transmitter. The first USRP-2940 is used as the Ma-MIMO GSM receiver. The second USRP-2940 device in the receiver is used to return the estimated CSI to the transmitter that has two USRPs using maximal ratio combining for the better CSI feedback. Least Square channel estimation method is used to obtain the CSI of the test environment. In order to reduce the channel estimation error in the feedback channel, feedback channel signals have been produced with an average of 30 dB SNR. The average SNR values in real time tests were obtained by using the equivalent signal power in order to compare with the simulation results. 

In this study, operating frequency for transmitter channels, operating frequency for feedback channels and bandwidth of devices are determined as 2.1 GHz, 2 GHz and 2 MHz, respectively. The carrier frequency offset and sample timing offset problems between the receiver and the transmitter for various reasons can be eliminated by the pilot sequences placed at certain frequencies to the transmitted signal.
Generally, GSM systems are designed with $N_u$ RF chains but we prefer to the different RF chains instead of the switching with the purpose of testing. The number of transmitted spatial bits is set to $3$ and $8$ different antenna subsets are created. As a single-carrier modulation method, $4$-QAM is used. The ML based and EDAS methods are compared in point of the impacts on the error performance.

\section{Results}
\begin{figure}[!b]
	\centering
	\includegraphics[width=\linewidth, height=4cm]{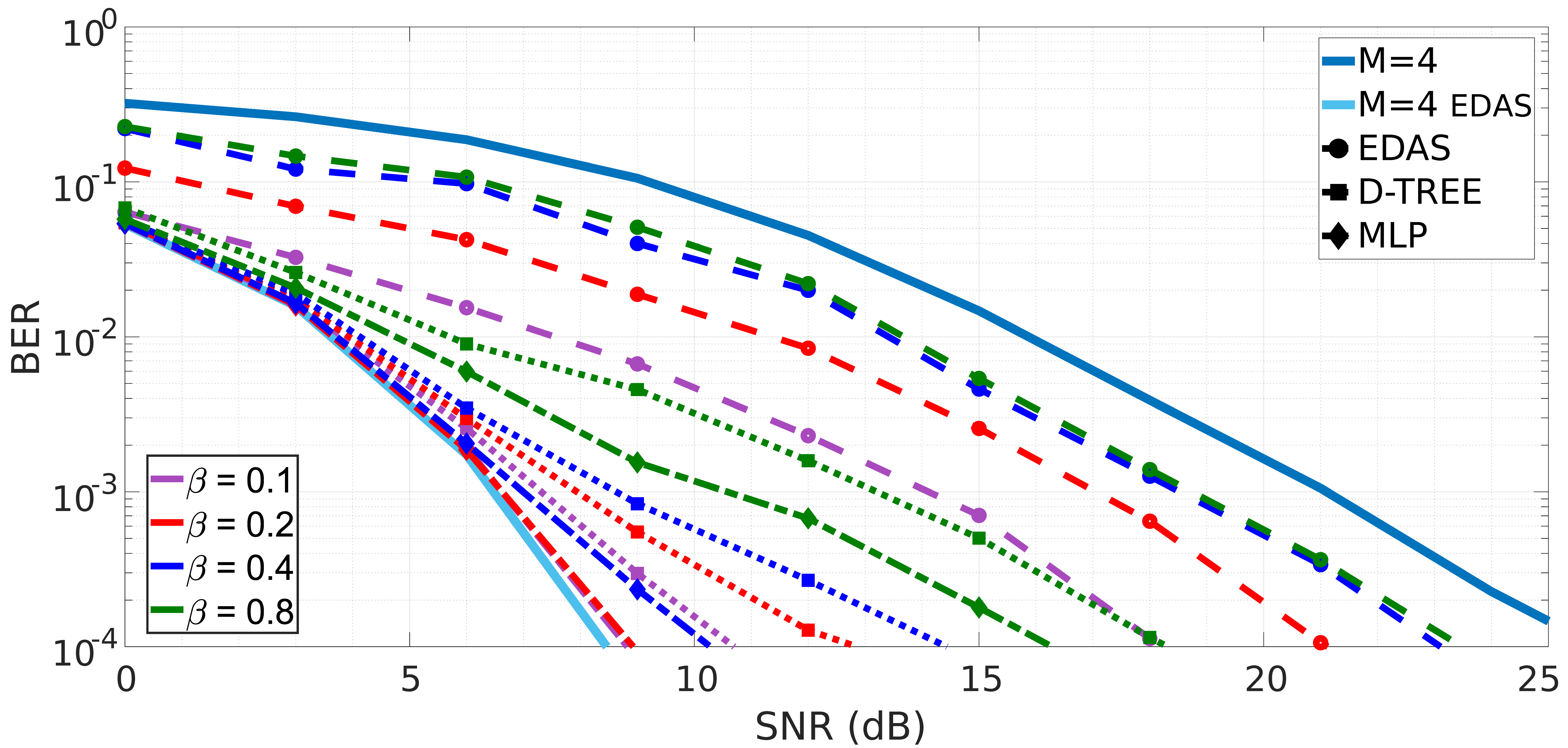}
	\caption{Simulation results for the \textit{i.i.d.} channel under the channel estimation errors.}
	\label{2a}
\end{figure}
In Ma-MIMO GSM system, the error performances of the proposed ML based antenna selection methods and EDAS are examined numerically and in the real-time environment. Both simulation and real-time system results are obtained, while the depth of the decision tree algorithm, the number of hidden layers of MLP and the number of neurons at one layer are equal to 17, 15, 10, respectively.
\begin{figure}[t!]
	\centering
	\includegraphics[width=\linewidth, height=4.15cm]{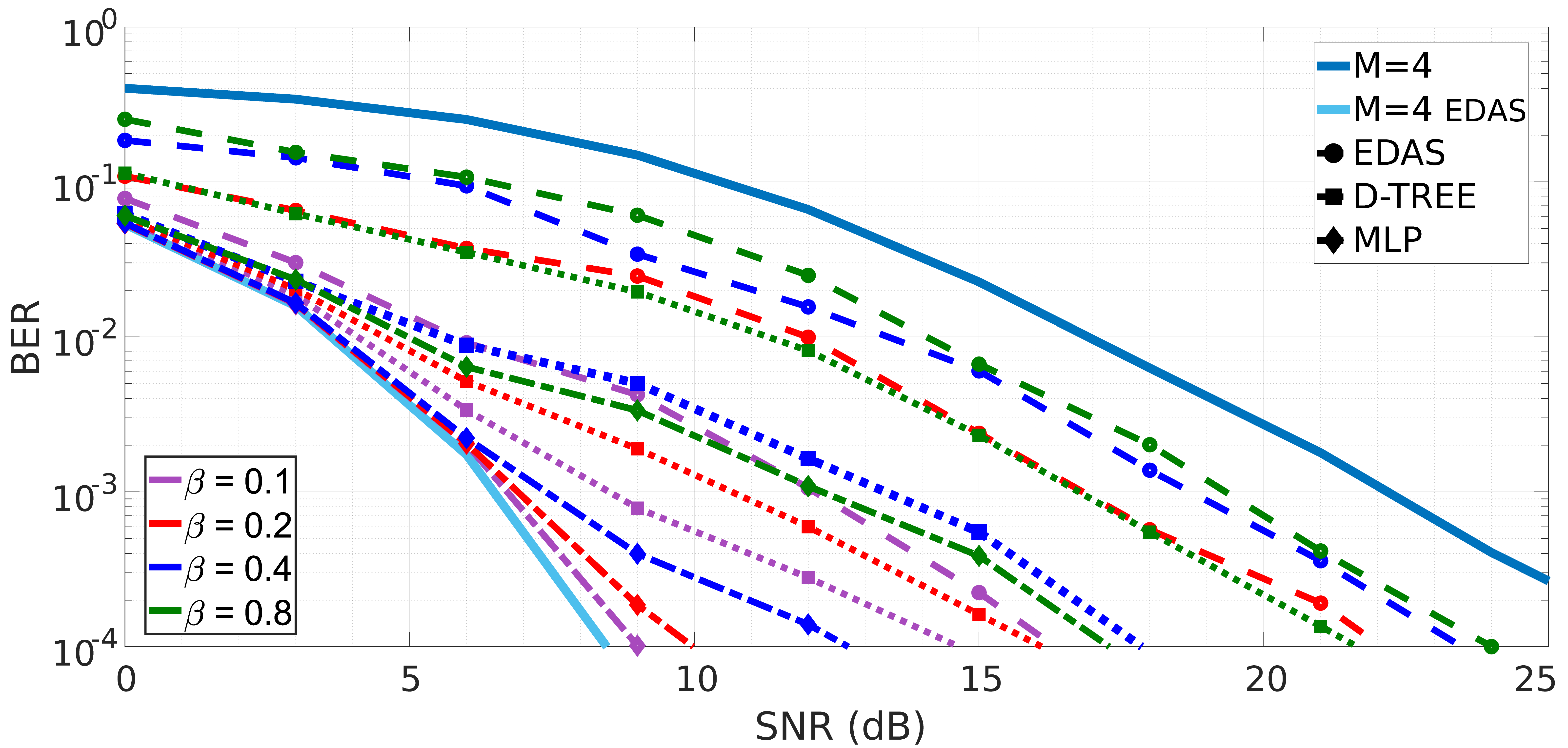}
	\caption{Simulation results for the time correlated channel and the channel estimation errors.}
	\label{2b}
\end{figure} 

In the simulation environment, the \textit{i.i.d.} and time-correlated channel models are evaluated for different estimation errors to simulate the real-time conditions. Time-correlated channel coefficient is set to $0.9$ and it is found that time-correlated channel has upper and lower boundaries similar to the \textit{i.i.d.} channel. However, in the presence of imperfect CSI, time-correlated channel appears to perform worse than the \textit{i.i.d.} channel. As shown in Figure \ref{2a} and Figure \ref{2b}, the value of the channel estimation error increases while the error performance of EDAS decreases. It is observed that ML performance is better than EDAS for the \textit{i.i.d.} channel. Performance of EDAS for the time-correlated channel is evaluated over the delayed channel structure as in real-time tests. In addition, EDAS provides a poor performance in time-correlated channels and under the presence of estimation errors.

\begin{figure}[t!]
	\centering
	\includegraphics[width=\linewidth, height=4.15cm]{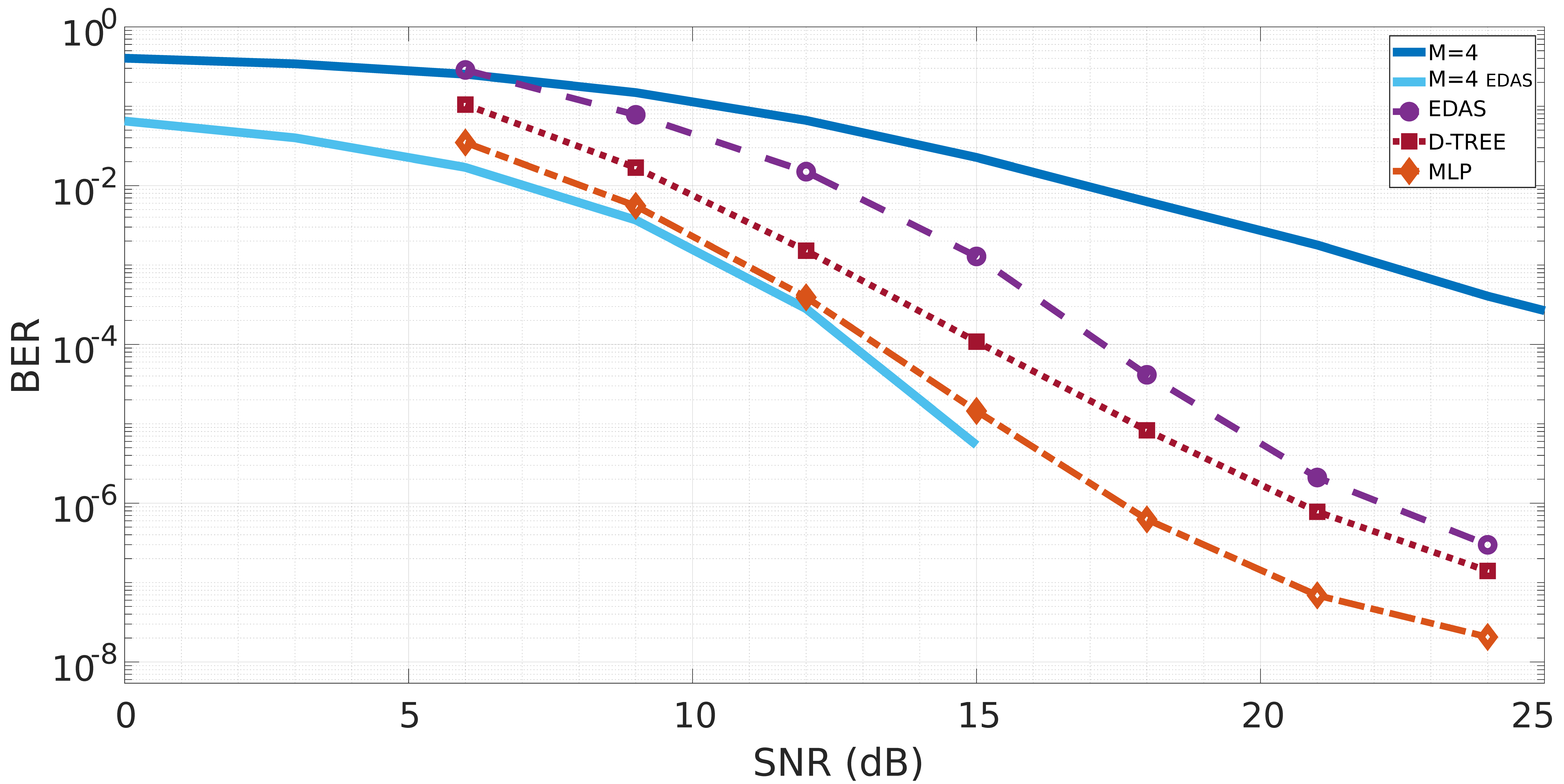}
	\caption{Real-time system results.}
	\label{2c}
\end{figure}

In the real-time environment tests, data gathering, training and test phases are limited to 120 seconds for each ML method. Collecting data duration is arranged 7 seconds and the same time intervals are applied for every 120 seconds. The results of the real-time tests are shown in Figure \ref{2c}. EDAS shows a weak performance due to the delayed imperfect CSI. EDAS performance increases and converges to the perfect CSI boundaries as the estimation error is reduced for the high SNR values. Although EDAS and decision-tree performances are similar for the high SNR values, decision-tree shows the higher performance compared with EDAS when SNR decreases. The other ML approach, MLP, achieves the lower error rates than decision tree but it is more complex method, which slows down the test system in real-time. As seen from Figure \ref{2c}, ML approaches provide well performance in presence of imperfect CSI and/or delayed CSI when compared to EDAS and decrease BER. Figures \ref{2a}, \ref{2b} and \ref{2c} show that the proposed ML based methods have a better performance when compared to the EDAS, in scenarios where channel estimation errors and time-correlated channel are present.\section{Conclusions}
The spectral efficiency and energy efficiency of the wireless systems can be improved using GSM technology. In this paper, GSM is implemented in a real-time test-bed for the first time in the literature. The performance of EDAS scheme is evaluated. In order to improve the performance in realistic deployment scenarios, mainly targeting correlated channels and the imperfect channel estimates, a machine learning based framework is proposed for the transmitter. Through simulation and test results, the error performances of the investigated schemes are compared, demonstrating the efficacy of the machine learning based approaches in the preference of real-life impairments.
\vspace{-2 mm}
\section*{Acknowledgment}
{\footnotesize This research was supported by TUBITAK under grant number 1160179.\par}
\bibliographystyle{IEEEtran}
\bibliography{ref}

\begin{thebibliography}{10}
\providecommand{\url}[1]{#1}
\csname url@samestyle\endcsname
\providecommand{\newblock}{\relax}
\providecommand{\bibinfo}[2]{#2}
\providecommand{\BIBentrySTDinterwordspacing}{\spaceskip=0pt\relax}
\providecommand{\BIBentryALTinterwordstretchfactor}{4}
\providecommand{\BIBentryALTinterwordspacing}{\spaceskip=\fontdimen2\font plus
\BIBentryALTinterwordstretchfactor\fontdimen3\font minus
  \fontdimen4\font\relax}
\providecommand{\BIBforeignlanguage}[2]{{%
\expandafter\ifx\csname l@#1\endcsname\relax
\typeout{** WARNING: IEEEtran.bst: No hyphenation pattern has been}%
\typeout{** loaded for the language `#1'. Using the pattern for}%
\typeout{** the default language instead.}%
\else
\language=\csname l@#1\endcsname
\fi
#2}}
\providecommand{\BIBdecl}{\relax}
\BIBdecl

\bibitem{S1}
P.~{Patcharamaneepakorn}, S.~{Wu}, C.~{Wang}, e.~M.~{Aggoune}, M.~M.
  {Alwakeel}, X.~{Ge}, and M.~D. {Renzo}, ``Spectral, energy, and economic
  efficiency of {5G} multicell massive {MIMO} systems with generalized spatial
  modulation,'' \emph{IEEE Trans. on Vehicular Tech.}, vol.~65, no.~12, pp.
  9715--9731, Dec 2016.

\bibitem{S2}
X.~{Gao}, O.~{Edfors}, F.~{Tufvesson}, and E.~G. {Larsson}, ``Massive {MIMO} in
  real propagation environments: Do all antennas contribute equally?''
  \emph{IEEE Trans. on Comm.}, vol.~63, no.~11, pp. 3917--3928, Nov 2015.

\bibitem{S12}
R.~{Rajashekar}, K.~V.~S. {Hari}, and L.~{Hanzo}, ``Antenna selection in
  spatial modulation systems,'' \emph{IEEE Comm. Letters}, vol.~17, no.~3, pp.
  521--524, March 2013.

\bibitem{S4}
J.~{Joung}, ``Machine learning-based antenna selection in {Wireless Comm.}''
  \emph{IEEE Comm. Letters}, vol.~20, no.~11, pp. 2241--2244, Nov 2016.

\bibitem{S5}
D.~{He}, C.~{Liu}, T.~Q.~S. {Quek}, and H.~{Wang}, ``Transmit antenna selection
  in {MIMO} wiretap channels: A machine learning approach,'' \emph{IEEE
  Wireless Comm. Letters}, vol.~7, no.~4, pp. 634--637, Aug 2018.

\bibitem{S6}
W.~{Qu}, M.~{Zhang}, X.~{Cheng}, and P.~{Ju}, ``Generalized spatial modulation
  with transmit antenna grouping for massive {MIMO},'' \emph{IEEE Access},
  vol.~5, pp. 26\,798--26\,807, 2017.

\bibitem{S7}
T.~{Datta} and A.~{Chockalingam}, ``On generalized spatial modulation,'' in
  \emph{IEEE WCNC}, April 2013, pp. 2716--2721.

\bibitem{S8}
A.~{Younis}, R.~{Mesleh}, M.~D. {Renzo}, and H.~{Haas}, ``Generalised spatial
  modulation for large-scale {MIMO},'' in \emph{EUSIPCO}, Sep. 2014, pp.
  346--350.

\bibitem{pillay2017improved}
N.~Pillay and H.~Xu, ``Improved generalized spatial modulation via antenna
  selection,'' \emph{International Journal of Comm. Systems}, vol.~30, no.~10,
  p. e3236, 2017.

\bibitem{S9}
B.~{Nosrat-Makouei}, J.~G. {Andrews}, and R.~W. {Heath}, ``{MIMO} interference
  alignment over correlated channels with imperfect {CSI},'' \emph{IEEE Trans.
  on Signal Processing}, vol.~59, no.~6, pp. 2783--2794, June 2011.

\bibitem{S13}
M.~{Sadek}, A.~{Tarighat}, and A.~H. {Sayed}, ``Exploiting spatio-temporal
  correlation for rate-efficient transmit beamforming,'' in \emph{Conference
  Record of the Thirty-Eighth Asilomar Conference on Signals, Systems and
  Computers, 2004.}, vol.~2, Nov 2004, pp. 2027--2031 Vol.2.

\bibitem{S10}
E.~Alpaydin, \emph{Introduction to Machine Learning}, 2nd~ed.\hskip 1em plus
  0.5em minus 0.4em\relax The MIT Press, 2010.

\bibitem{S11}
A.~G{\'{e}}ron, \emph{{Hands-On Machine Learning with Scikit-Learn and
  TensorFlow}}.\hskip 1em plus 0.5em minus 0.4em\relax O'Reilly Media, 2017.

\end{thebibliography}

\end{document}